\documentclass[twocolumn, 10pt,
superscriptaddress,
 amsmath,amssymb,
 aps,
 prl,
floatfix,
]{revtex4-2}

\preprint{APS/123-QED}
\usepackage{graphicx}
\usepackage{dcolumn}
\usepackage{adjustbox}
\usepackage{bm}
\usepackage{multirow}
\usepackage{xcolor}

\begin{document}
\title{Raman-enabled platicon microcomb in 4H-SiC microresonator}
\author{Jingwei Li}
\author{Ruixuan Wang}
\author{Qing Li}
\email{qingli2@andrew.cmu.edu}
\affiliation{Department of Electrical and Computer Engineering, Carnegie Mellon University, Pittsburgh, PA 15213, USA}


\begin{abstract}
Stimulated Raman scattering in a Kerr microresonator is generally considered a competing nonlinear process that hinders the formation of Kerr soliton microcombs. In this work, we experimentally demonstrate that the ubiquitous Raman gain in Kerr microresonators can, in fact, be harnessed to achieve the opposite effect: it enables the formation of platicon microcomb in the normal dispersion regime, while also relaxing the conditions for soliton formation and broadening the spectrum through the simultaneous excitation of a Stokes soliton. We showcase this process in a compact silicon carbide microresonator supporting a platicon microcomb spanning 1500 to 1700 nm, with a pump-to-comb conversion efficiency as high as $56\%$. Furthermore, the presence of a Stokes soliton in the same mode family extends the comb spectrum beyond 1800 nm. By intentionally leveraging—rather than suppressing—the Raman effect, our work offers new insights into the Raman–Kerr interplay and introduces a promising approach to generating broadband platicon microcombs.
\end{abstract}
\maketitle
\noindent

 \textit{Introduction}-Microresonator-based soliton frequency combs, or microcombs, have transformed optical technology across a broad spectrum of applications, including precision metrology \cite{Diddams_comb_review1, Papp_comb_synthesizer, Comb_clock1}, imaging and sensing \cite{Vahala_comb_imaging, Diddams_comb_sensing1, Comb_midIR_spectroscopy, Vahala19_comb_exoplanets}, light detection and ranging \cite{Vahala_comb_ranging, Kippenberg_comb_ranging}, optical communication \cite{Moss_comb_communication_review}, microwave photonics \cite{Yi_microwave_nature, Gaeta_microwave_nature, Diddams_microwave_Nature}, as well as quantum information \cite{Review_quantum_comb, Vuckovic_4HSiC_soliton, XuYi_squeezed_comb}. The most widely adopted method for generating soliton microcombs leverages the Kerr nonlinearity in various integrated photonic materials such as  silicon \cite{Gaeta_Si_comb, Gaeta_Si_dual_midIR}, silica \cite{Vahala_comb_silica, Xiao_comb_silica}, aluminum nitride \cite{Guo_comb_AlN, Tang_comb_AlN_ref}, silicon nitride \cite{Xue_dark_soliton, Li_SiN_octave,Kippenberg_SiN_octave, Wong_soliton_aux, Gaeta_comb_battery}, silicon carbide \cite{Vuckovic_4HSiC_soliton, OuXin_soliton}, lithium niobate \cite{Loncar_LN_EOM, Lin_LN_soliton, Tang_LN_comb, Lin_LN_soliton, Yang_dark_pulse_LN}, AlGaAs \cite{Bowers_comb_AlGaAs}, etc. In the time domain, Kerr soliton microcombs manifest mainly as bright or dark pulses, depending on whether the microresonator operates in the anomalous or normal dispersion regime \cite{Chembo_comb_stability}, respectively. Compared to bright solitons, dark solitons offer higher pump-to-comb conversion efficiency and enable comb generation in the normal dispersion regime—both advantageous for practical deployment \cite{Oxenlowe_Petabit_dark_comb, Diddams_microwave_Nature}. 

Despite their advantages, dark pulse solitons in Kerr microresonators present unique challenges. A key limitation is their typically narrow spectral bandwidth \cite{Maleki_normal_GVD,Xue_dark_soliton, TorresC_dark_solitons, Yang_dark_pulse_LN, Diddams_microwave_Nature}. Moreover, accessing the dark soliton state with a continuous-wave pump often requires introducing local dispersion perturbations-either through modal interactions within a single resonator \cite{Xue_dark_soliton} or by coupling adjacent resonators \cite{TorresC_dark_solitons}-which constrains the soliton accessibility range. In this work, we demonstrate that platicon, a high-order periodic dark soliton featuring unique flattop shape \cite{Raman_Kerr_theory, Wong_platicon_stimulated, Kippenberg_platicon_2022}, can be reliably generated via a non-local perturbation arising from the Raman effect in a crystalline Kerr microresonator. (Note platicon is also considered as a bright pulse in some literature since for periodic pulses, the difference between bright and dark ones disappears \cite{Raman_Kerr_theory}.) In addition, the spectral bandwidth of platicon can be significantly enhanced through the excitation of a coexisting Stokes soliton. Our result differs from prior Stokes soliton demonstrations in amorphous materials such as silica \cite{Vahala_Stokes_soliton, Yao_Stokes_sensing}, where the Kerr microcomb is first generated from the anomalous dispersion and the Stokes soliton is subsequently excited in a different mode family. In contrast, the Stokes signal in our experiment is first generated by a pump laser in the normal dispersion regime, and the resultant platicon and the Stokes soliton interact within the same mode family. On the other hand, although several numerical studies have predicted Raman-assisted platicon generation in the normal dispersion regime \cite{RamanKerr_theory_2017,Tlidi_RamanKerr_theory2021,Zhao_stokes_Kerr, Su_Raman_dark_soliton, Wang_Kerr_Platicon}, to the best of our knowledge, experimental validation in integrated Kerr microresonators has remained elusive until now \cite{Chen_Raman_normal_FP}. In fact, most prior work has sought to suppress the Raman effect to enable Kerr comb generation \cite{Gaeta_Raman_competition, Loncar_LN_octave_soliton, Yang_dark_pulse_LN}. By instead harnessing the intrinsic Raman gain, our approach challenges this prevailing strategy, offering new insights into Raman–Kerr dynamics and paving the way for broadband platicon microcombs.

\begin{figure*}[ht]
\centering
\includegraphics[width=0.9\linewidth]{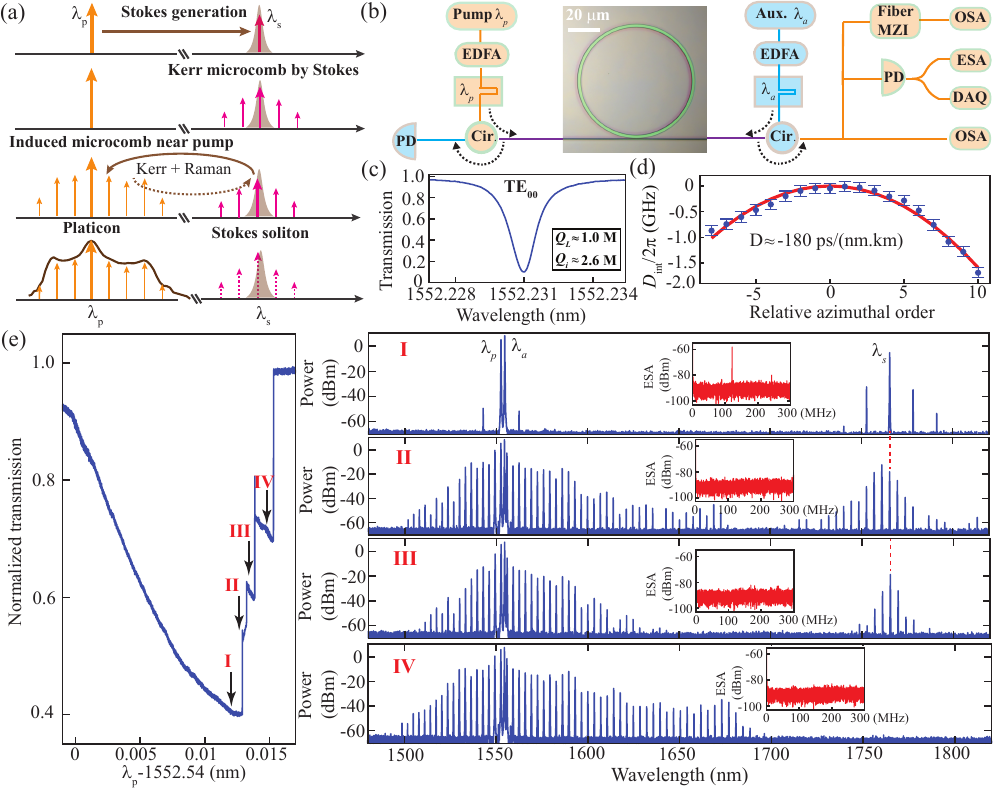}
\caption{(a) Illustration of the Raman-Kerr interaction in a crystalline microresonator with normal dispersion near the pump wavelength ($\lambda_p$). The shaded area near the Stokes line ($\lambda_s$) depicts the Raman gain profile. (b) Experimental schematic for a 43-$\mu$m-radius 4H-silicon-carbide (4H-SiC) microresonator where an auxiliary pump laser ($\lambda_a$) is introduced to mitigate the thermal effect. EDFA: erbium-doped fiber amplifier; Cir: circulator; PD: photodetector; DAQ: data acquisition; MZI: Mach-Zehnder interferometer; ESA: electrical spectrum analyzer; and OSA: optical spectrum analyzer. 100-GHz bandpass filters are employed for the pump and auxiliary lasers to remove the noise from EDFAs. (c) Linear swept-wavelength transmission of the fundamental transverse-electric (TE$_{00}$) mode around 1552 nm for a microring with radius of $43.72\ \mu$m and a width of $2.1\ \mu$m. (d) Experimentally measured group-velocity dispersion (GVD, blue markers) and its linear fitting (red line), confirming normal GVD in the 1550 nm band. (e) The left panel is the transmission scan of the 1552 nm resonance shown in (c) at an on-chip pump power of 50 mW ($\lambda_a \approx 1554.6$ nm). The right panel displays the optical spectra corresponding to the marked detunings (I to IV) on the left, with the insets plotting the photodetected electrical spectrum (resolution bandwidth of 100 kHz).}
\label{Fig1}
\end{figure*}

\textit{Results}-We investigate the generation of Raman-Kerr-induced platicon in compact 4H-silicon carbide (4H-SiC) microring resonators with a radius of 43 $\mu$m. Crystalline 4H-SiC exhibits a dominant Raman shift of 777 cm$^{-1}$ ($23.3$ THz) and a narrow gain bandwidth of 4 cm$^{-1}$ (120 GHz) \cite{Li_SiC_Raman}, which is smaller than the free spectral range (FSR) of the microresonator of approximately 400 GHz. For a SiC thickness of 720 nm, the fundamental transverse-electric (TE$_{00}$) mode is expected to possess normal dispersion in the 1550 nm band when the ring width exceeds $1.6$ $\mu$m (see Appendix). A schematic of the envisioned Raman-Kerr interaction is shown in Fig.~1(a). First, a Stokes signal is generated through stimulated Raman scattering by pumping one of the TE$_{00}$ resonances. Next, a Kerr microcomb forms near the Stokes wavelength if local anomalous dispersion is present. Then, a four-wave mixing Bragg scattering (FWM-BS) process occurs, in which the pump laser and Stokes signal beat together, transferring comb lines near the Stokes wavelength back to the pump region and hence stimulating the formation of the platicon microcomb \cite{Li_SiC_Raman}. Finally, a Stokes soliton may also be excited, with a spectrum that may or may not overlap with that of platicon.

To mitigate the thermo-optic bistability that renders the soliton state thermally unstable, we employ an auxiliary laser to stabilize the cavity temperature during soliton access in the TE$_{00}$ resonance (Fig.~1(b))\cite{Wong_soliton_aux, Li_4HSiC_100GHz_soliton}. The fundamental transverse-magnetic (TM$_{00}$) resonance is chosen for auxiliary laser cooling due to its strong over-coupling in the 1550 nm band (see Appendix), which helps suppress competing nonlinear effects. In contrast, the TE$_{00}$ resonance is found to be near the critical coupling condition (see Fig.~1(c) for an example with a $43.72$-$\mu$m radius and $2.1$-$\mu$m width) and exhibits normal dispersion in the 1550 nm band (Fig.~1(d)).

\begin{figure*}[ht]
\centering
\includegraphics[width=0.9\linewidth]{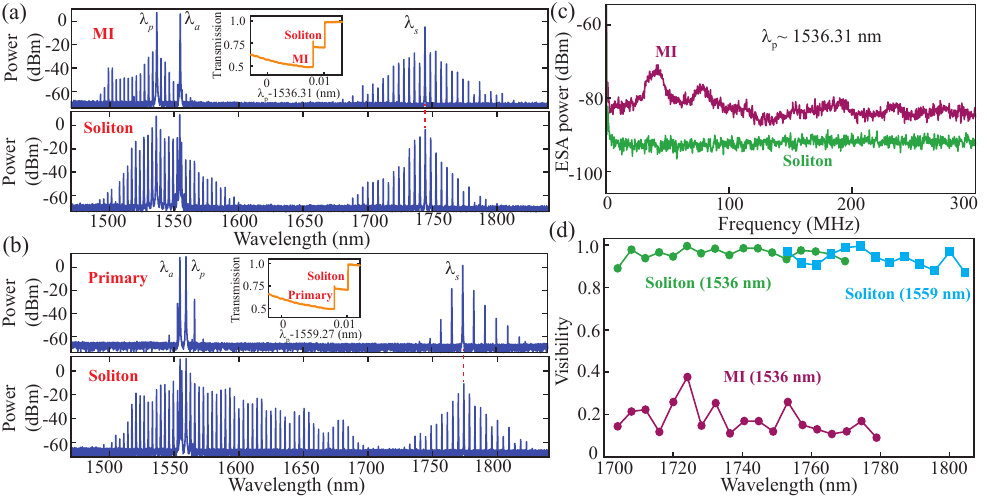}
\caption{(a) Optical spectra corresponding to the modulation instability (MI) and the single soliton step (see inset) of the same microring in Fig.~1(e) when pumped at $1536.3$ nm and a pump power of 50 mW. (b) Optical spectra corresponding to the primary comb and the single soliton step (see inset) when pumped at $1559.3$ nm and a pump power of 50 mW. (c) Superimposed ESA spectra of the photodetector corresponding to the MI and soliton states in (a). (d) Visibility measurement results by sending the optical output to an asymmetric fiber MZI and recording multiple optical spectra to compute the contrast (see Fig.~1(b)).}
\label{Fig2}
\end{figure*}

When the on-chip power is increased to 50 mW for the TE$_{00}$ resonance in Fig.~1(c), several distinct soliton steps appear in the pump laser transmission scan (see Fig.~1(e)). For instance, when the pump is held near the bottom of the transmission curve (detuning \textbf{I} in Fig.~1(e)), the optical spectrum is dominated by the pump laser ($\lambda_p\approx 1552.53$ nm), the auxiliary laser ($\lambda_a\approx1554.6$ nm), and the Stokes signal ($\lambda_s \approx 1765.51$ nm). The observed Raman shift is 777 cm$^{-1}$, matching the center of the Raman gain profile \cite{Li_SiC_Raman}. Additional comb lines near the Stokes arise from parametric gain enabled by the resonator's weak anomalous dispersion around 1760 nm (see Appendix). These comb lines are further transferred to the 1550 nm region via FWM-BS, albeit at significantly lower power levels. As the pump is slowly tuned into the first soliton step (detuning \textbf{II}), a comb profile characteristic of platicon-featuring two pronounced wings-emerges in the 1550 nm band. Concurrently, multiple comb lines appear around the Stokes wavelength, with the peak comb line blue-shifted by one FSR. In the following soliton step (detuning \textbf{III}), the comb spans around both the pump and the Stokes shrink significantly, and the Stokes-region peak shifts back to its original position. Finally, in the last soliton step (detuning \textbf{IV}), the comb around the Stokes disappears entirely, leaving behind the platicon microcomb that spans from 1500 nm to 1700 nm ($\approx 23$ THz). The observed spectrum of the platicon has a reasonable agreement with numerical simulation (see Appendix), with its pump-to-comb power efficiency estimated at $56\%$.

The coherence properties of the various comb states shown in Fig.~1(e) are examined by analyzing their photodetected transmission using an electrical spectrum analyzer (ESA), as illustrated in Fig.~1(b). At the bottom of the pump transmission curve (detuning \textbf{I}), the ESA spectrum displays at least one prominent peak within the measured frequency range ($0.1$-300 MHz), indicating a noisy state-likely a breather \cite{Chembo_comb_stability, Zhao_stokes_Kerr, Wang_Kerr_Platicon}. In contrast, the ESA spectra for detunings \textbf{II} through \textbf{IV} exhibit low noise levels comparable to the detector’s dark noise (see Appendix), suggesting high coherence.

When the on-chip power is further increased from 50 mW to 80 mW, similar soliton steps are observed in the pump transmission (see Fig.~\ref{FigE_beatnote} in Appendix). However, the ESA spectra at detunings \textbf{II} and \textbf{III} now show increased noise, in contrast to their low-power counterparts in Fig.~1(e). This change implies that at higher powers, the interplay between Raman and Kerr effects can indeed destabilize soliton formation \cite{Gaeta_Raman_competition}. Remarkably, the final soliton step (detuning \textbf{IV}), which corresponds to a platicon without the Stokes soliton, continues to exhibit low phase noise even at this elevated power. Its coherence is further confirmed by selecting a comb line and measuring its beat note with an external tunable laser (see Fig.~\ref{FigE_beatnote}(c) in Appendix), which shows similar behaviors as the ESA data.

\begin{figure*}[ht]
\centering
\includegraphics[width=0.9\linewidth]{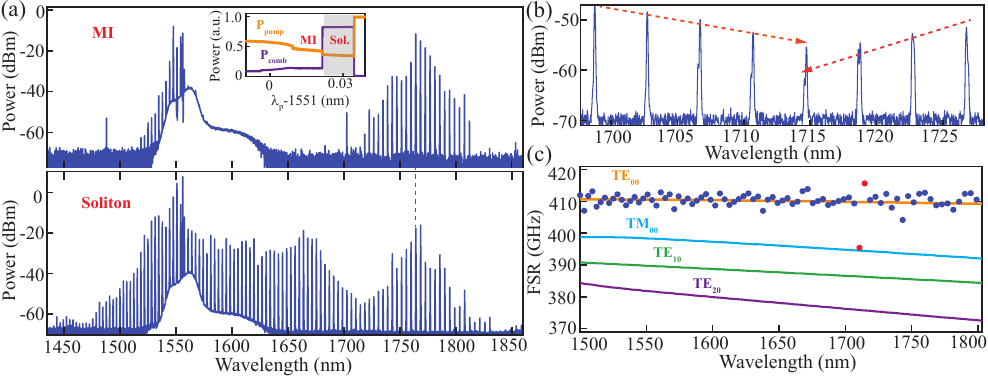}
\caption{(a) Optical spectra corresponding to the MI and soliton state (see inset) of a 4H-SiC microring with radius of $43.56\ \mu$m at an on-chip pump power of 100 mW. For the MI comb spectrum, the pump and auxiliary lasers are already notched out, while for the soliton spectrum all the filters are removed. (b) Zoomed-in spectrum near the wavelengths of 1700 nm. (c) Extracted frequency spacing of soliton comb lines from the OSA (blue markers) and its comparison against simulation results (solid lines) for various resonant mode families.}
\label{Fig3}
\end{figure*}

In previous demonstrations of dark pulse solitons, only a limited number of resonances could support soliton formation, as it critically depended on the presence of local anomalous dispersion \cite{Xue_dark_soliton, TorresC_dark_solitons}. In contrast, the platicon microcomb observed in Figs.~1(e) is enabled by non-local perturbation from the ubiquitous Raman gain, suggesting a wider range of accessible pump wavelengths. To validate this hypothesis, we tested several distinct pump resonances for soliton generation. For instance, Fig.~2(a) shows experimental results for the $1536.3$ nm resonance, where a single soliton step is observed in the pump transmission. The corresponding optical spectrum includes a coexisting Stokes comb. Similar behavior is found in other resonances, such as the 1559 nm mode shown in Fig.~2(b). ESA measurements distinguish the noise characteristics between MI and soliton states, as illustrated in Fig.~2(c).

Direct beat note measurements near the Stokes wavelength are limited by the unavailability of tunable lasers in the 1800 nm range. As an alternative, we assess phase noise by passing the comb lines through a fiber-based asymmetric Mach-Zehnder interferometer (MZI) with a 3-meter path length difference (corresponding to a 68 MHz FSR). For a coherent source such as a laser, thermally induced phase fluctuations in the fiber cause the MZI output intensity to vary between zero and its maximum. In contrast, input light with high intrinsic phase noise produces significantly smaller output fluctuations. Using this principle, we capture more than 100 traces on an OSA (Yokogawa 6375E) to determine the visibility of each comb line \cite{Gaeta_tworing_darksoliton}. The high visibility measured for the soliton states in Figs.~2(a) and 2(b) confirms their low phase noise, in contrast to the low visibility associated with the noisy MI state at the $1536.3$ nm pump (see Fig.~2(d)).

The results in Fig.~1(e) suggest that strong competition between Raman and Kerr effects can lead to the annihilation of the Stokes soliton, making platicon the most stable solution in the system. In our SiC chip, which contains more than 50 microrings, this behavior is observed only in a few devices that exhibit the strongest Raman gain-specifically, those where the Stokes wavelength is well aligned with the center of the Raman gain profile. In most other devices, the coexistence of platicon and a Stokes soliton is more common. For instance, Fig.~3 presents results from a different SiC microring with a radius of $43.56$ $\mu$m. This device shows a Raman shift of $779.8$ cm$^{-1}$ ($23.4$ THz), indicating reduced Raman gain relative to the center value of 777 cm$^{-1}$. As shown in Fig.~3(a), a single soliton step appears in the pump transmission following the MI stage, accompanied by a notable increase in comb power. The pump-to-comb conversion efficiency is estimated at approximately $30\%$, measured by filtering out the pump and auxiliary lasers prior to photodetection. 

Within the soliton step, the spectra of the platicon and Stokes soliton overlap. A zoomed-in view around 1700 nm (Fig.~3(b)) reveals that their absolute frequencies are not perfectly aligned-an expected feature of synchronized soliton pulses \cite{Vahala_Stokes_soliton}. Using fine OSA scans with a wavelength resolution of 50 pm, we extract the FSRs of the Kerr and Stokes solitons in Fig.~3(c). Comparing these experimentally measured FSRs with simulated results for various resonant modes confirms that both solitons are generated within the same TE$_{00}$ mode family. The maximum wavelength splitting occurs near 1715 nm (marked by two red dots in Fig.~3(c)) and is approximately 150 pm, corresponding to a frequency difference of 15 GHz.

\textit{Conclusion}-We have experimentally demonstrated the generation of platicon microcombs in compact 4H-SiC microresonators by leveraging non-local perturbation from the intrinsic Raman effect-an approach distinct from previous methods that relied on local dispersion engineering through modal interactions in single or coupled resonators. This technique significantly broadens the soliton access window, while the simultaneous generation of a Stokes soliton further extends the spectral bandwidth. In 43-$\mu$m-radius 4H-SiC microrings, the combined spectral span of platicon and the Stokes soliton reaches from 1450 nm to 1850 nm, thanks to its large Raman shift. Our findings also reveal the complex dynamics arising from the interplay between Raman and Kerr nonlinearities in the normal dispersion regime. While strong competition between these effects can suppress the Stokes soliton, a more balanced interaction supports their coexistence. These results offer new insight into Raman-Kerr interactions in single-crystalline microresonators and highlight a promising pathway for achieving broadband, power-efficient Kerr and Stokes soliton generation in systems with normal dispersion.

\textit{Acknowledgments}-This work was supported by NSF (2240420, 2131162, 2427228) and CableLabs University Outreach Program. The authors acknowledge helpful discussions with Prof.~Haiyan Ou from DTU, the use of Bertucci Nanotechnology Laboratory at Carnegie Mellon University supported by grant BNL-78657879, and the Materials Characterization Facility supported by grant MCF-677785. R.~Wang and J.~Li also acknowledge the support of Tan Endowed Graduate Fellowship and Benjamin Garver Lamme/Westinghouse Graduate Fellowship from CMU, respectively. 

\textit{Data Availability}-Data underlying the results presented in this paper are available upon reasonable request.

\section*{Appendix}

\textbf{Device fabrication and linear properties}-The device fabrication begins with a 4H-SiC-on-insulator chip consisting of an approximate 720 nm silicon 4H-SiC layer on top of a 2 $\mu$m oxide layer (NGK Insulators) \cite{Li_4HSiC_comb}. The device pattern is initially defined by spin-coating a 1-$\mu$m-thick negative electron-beam resist (flowable oxide, FOx-16), which serves as the etching mask. This pattern is then written using a 100 kV electron-beam lithography system (Elionix ELS-G100). After development, the pattern is transferred into the SiC layer using a CHF$_3$/O$_2$ plasma etching process, achieving an etch depth of approximately 600 nm and leaving around 120 nm of unetched SiC. Subsequently, a 2 $\mu$m oxide cladding is deposited on top of the 4H-SiC layer to encapsulate the devices. Broadband coupling is achieved by employing lensed fibers and implementing inverse tapers (with taper width of 250 nm) on the SiC chip \cite{Li_SiC_Raman}. The total insertion loss is measured to be around 13-17 dB for various rings \cite{Li_SiC_Raman}.

\begin{figure}[ht]
\centering
\includegraphics[width=1.0\linewidth]{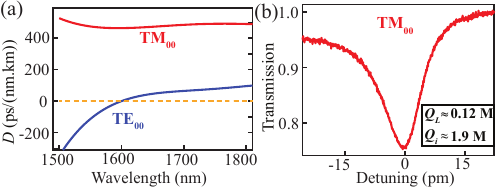}
\caption{(a) Simulated group-velocity dispersion for the TE$_{00}$ and TM$_{00}$ resonances in a $43$-$\mu$m-radius SiC microring. (b) Representative TM$_{00}$ resonance measured in the 1550 nm band, which is strongly over-coupled.}
\label{FigE_linear}
\end{figure}

In Fig.~\ref{FigE_linear}(a), the simulated dispersion profile of the TE$_{00}$ and TM$_{00}$ resonances are provided. As can be seen, the TE$_{00}$ mode family exhibits normal dispersion in the 1550 nm band, which becomes slightly anomalous in the longer wavelength range ($>1600$ nm). In contrast, the TM$_{00}$ exhibits strong anomalous dispersion throughout the entire wavelength range of interest, but its Kerr effect is suppressed due to the resonance being strongly over-coupled (see Fig.~\ref{FigE_linear}(b)). As such, the auxiliary laser can couple to the TM$_{00}$ resonance without inducing any significant nonlinear effects.

\begin{figure}[ht]
\centering
\includegraphics[width=1.0\linewidth]{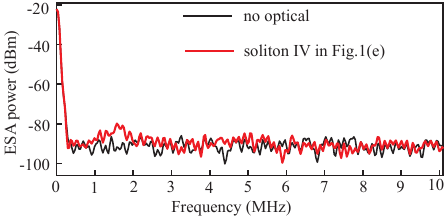}
\caption{Zoomed-in ESA spectra in the frequency range of $0.1$-10 MHz for the photodetector used in this work: the black line corresponds to zero optical input (dark noise) and the red line corresponds to platicon in Fig.~1(e) in the main text. The resolution bandwidth of the ESA is set at 100 kHz. }
\label{FigE_PD}
\end{figure}

\textbf{ESA noise benchmarking}-The photodetector used in this work is Newport InGaAs Nanosecond 1623 Detector which has a 3-dB bandwidth of 500 MHz and a conversion gain of 50 V/W. The dark noise of the photodetector is provided in Fig.~\ref{FigE_PD} (black line) for the frequency range of $0.1$-10 MHz, which is also compared against the ESA spectrum of the soliton state (state IV) in Fig.~1(e) in the main text (red line). Their almost identical spectral response confirms the low phase noise of the soliton state.

\begin{figure*}[ht]
\centering
\includegraphics[width=0.9\linewidth]{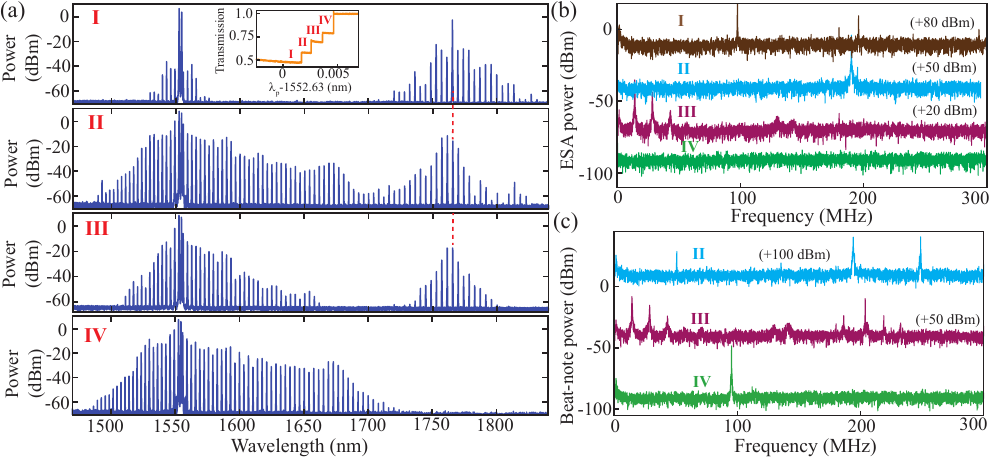}
\caption{(a) Optical spectra of the same resonance in Fig.~1(e) but at an increased on-chip pump power of 80 mW (the inset is the corresponding pump transmission). (b) ESA spectrum of the photodetector corresponding to the four different states marked in (a). (c) Beat note measurement for detunings from detunings \textbf{II} to \textbf{IV} by interfering the comb line near 1543 nm with an external laser tuned to similar wavelengths (the comb lines in the state \textbf{I} are too weak for this measurement).}
\label{FigE_beatnote}
\end{figure*}

\textbf{Platicon at increased pump powers}-In Fig.~\ref{FigE_beatnote}, we provide additional data on the pump transmission, optical spectra, and phase noise characterization for the $1552.3$ nm resonance of the $43.72$-$\mu$m-radius SiC microring (Fig.~1(c)) when pumped at 80 mW.

\begin{figure*}[ht]
\centering
\includegraphics[width=0.9\linewidth]{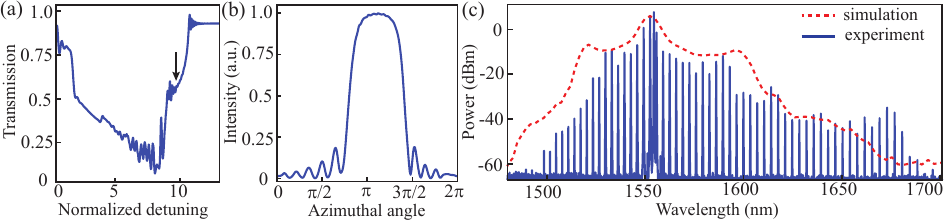}
\caption{(a) Simulated pump transmission for the device shown in Fig.~1(e), with the normalized detuning defined as the wavelength detuning normalized by the cavity linewidth. The vertical arrow indicates the soliton step; (b) Corresponding platicon pulse inside the microresonator; (c) Comparison of the optical spectrum between the experimental data (blue solid line) and simulation (red dashed line). In the simulation, we assumed a flat loaded and intrinsic $Q$ of $1.0$ million and $2.0$ million, respectively; a pump power of 50 mW; the Kerr nonlinear parameter $\gamma \approx 4\ \text{W}^{-1}\text{m}^{-1}$, and a peak Raman gain coefficient $g_R \approx 0.75 \ \text{cm}/\text{GW}$ \cite{Li_SiC_Raman}. In addition, the second- and third-order dispersion are set as $\beta_2=0.19\ \text{ps}^2/\text{m}$ and $\beta_3=2.5\times 10^{-3}\ \text{ps}^3/\text{m}$ at 1550 nm to match the normal dispersion of $D=-150 \ \text{ps}/\left(\text{nm}\cdot\text{km}\right)$ at 1550 nm and weak anomalous dispersion of $D=100 \ \text{ps}/\left(\text{nm}\cdot\text{km}\right)$ at 1760 nm.}
\label{FigE_Simulation}
\end{figure*}

\textbf{Numerical simulation}-To numerically study the Raman and Kerr interaction in SiC microresonators, we modified the Lugiato-Lefever equation (LLE) to incorporate the Raman gain \cite{Raman_Kerr_theory, Wang_Kerr_Platicon}. One such example is provided in Fig.~\ref{FigE_Simulation} based on device parameters similar to Fig.~1(e). As can be seen, reasonable agreement has been achieved between the simulation and experiment, with the main difference attributed to several simplifying assumptions in the numerical model, including dispersion (only second- and third-order dispersion terms are considered) and coupling (flat intrinsic and coupling quality factors). Modeling of more realistic device parameters and the replication of coexisting platicon and Stokes soliton is left to future work.


\bibliographystyle{apsrev4-1}
\bibliography{SiC_Ref}

\end{document}